\documentstyle[12pt]{article}
\title{Anomalies and Deconfinement \thanks {Talk at the IV Meeting on
Quantum Mechanics of Fundamental Systems, Santiago de Chile, December
27-30,1991}}
\vspace{3.5in}
\author{ Fidel A.Schaposnik \thanks{Investigador CICBA}\\
Department of Physics\\
University of Illinois at Urbana-Champaign\\
1110 W.Green St., Urbana, IL 61801, USA\\
and\\
Departamento de F\'\i sica, Universidad Nacional de La Plata \thanks{Permanent
Address}\\
C.C.67, 1900 La Plata, Argentina}
\date{}
\begin{document}
\newcommand{\beq} {\begin{equation}}
\newcommand{\eeq} {\end{equation}}
\newcommand{\ep} {\epsilon^{\mu\nu}}
\maketitle
\begin{abstract}

I discuss how instanton effects can be wiped-out due to the existence of
anomalies. I first consider Compact Quantum Electrodynamics in 3 dimensions
where confinement of electric charge is destroyed when fermions are
added so that a Chern-Simons term is generated as a one-loop effect. I
also show that a similar phenomenon occurs in the two-dimensional abelian
chiral Higgs model. In both cases anomalies (parity anomaly, gauge anomaly)
are responsible of the deconfinement mechanism.
\end{abstract}
\newpage

\section{Introduction}
Compact Quantum Electrodynamics in $d=2+1$ dimensions ($CQED_3$) is a nice
example on how non-perturbative effects can lead to the confinement of a
fundamental quantum number such as electric charge.

Consider for example a $d=(3+1)$ $SO(3)$ gauge theory with spontaneous symmetry
breaking through a Higgs field in the adjoint. As it is well-known,
the resulting classical
equations of motion  have regular {\it static} solutions: the well-honored
't Hooft-Polyakov monopoles \cite{tH}-\cite{Po1}. These static solutions
in $3+1$ dimensions can be taken as instanton solutions in  3-dimensional
Euclidean space-time. One can then analyse the effects of instantons in
the dimensionally reduced theory, 3-dimensional Compact Quantum
Electrodynamics (After symmetry breaking of the original
(compact) $SO(3)$ group, one ends with a residual compact abelian ($U(1)$)
symmetry ensuring the existence of {\it regular} solutions with finite
action).

In $d=2+1$ the Coulomb potential for a $U(1)$ gauge  theory is, at the
classical level, logarithmic. In the case of $CQED_3$, Polyakov \cite{Po2}
showed, fifteen years ago that the interaction between electric charges
becomes linear due to instanton (monopole) effects.

In fact, Polyakov \cite{Po2} observed that, at the one-loop level,
the theory is equivalent to a
Coulomb gas which exhibits Debye screening of magnetic (monopole) charges
for aribitrarily weak coupling. He then showed how this screening of
magnetic charge implies confinement of electric charge: due to the monopole
background, an electric string between two fixed charges is stabilized.
The length scale of the
crossover from logarithmic to linear potential becomes exponentially
large at weak coupling.

What happens with confinement when a Chern-Simons (CS) term is added
to this $d=3$ model? I shall address to this question in the first part of
my talk but, before advancing the
answer, let me discuss why this is an interesting question.

Originally \cite{DJT} the Chern-Simons term was introduced in $d=3$ gauge
theories
as a way of producing symmetry breaking without use of the Higgs fields:
due to the presence of the Chern-Simons term, the originally massless $QED_3$
photon becomes massive. For this reason the model is known as Topologically
massive $QED_3$.

Another important property of a gauge theory with Chern-Simons term can
be understood by analysing the classical equations of motion.
Calling $Q$ the electric charge and $\Phi$
the magnetic flux, one of the equations of motion leads to a basic
relation:

\beq
Q = \mu \Phi \label{A}
\eeq
where $\mu$ is the coefficient of the Chern-Simons term. If one
then considers a tube with magnetic flux (a Nielsen-Olesen vortex \cite{NO})
it necessarily carries electric charge \cite{PK}. Moreover, in the non-abelian
case $\mu$ needs to be quantized (due to gauge invariance
requirements \cite{DJT}) and then both flux and charge of vortices are
quantized \cite{dVS}.
One is then dealing with fractional statistic objects
a fact that attracted people from Condensed Matter
physics to the subject \cite{Ed}.

Another reason to include a Chern-Simons term when studying $QED_3$ or $QCD_3$
is the following: in $d=3$, integration over fermions leads to a Chern-
Simons term as a one-loop efect related to violation of parity invariance
\cite{R}.
In the same way a photon mass arises in $QED_2$ (Schwinger mechanism)
as a result of the impossibility of regularizing the fermion determinant both
in
a gauge invariant and a chiral invariant way, the impossibility of regularizing
the $d=3$ fermion determinant both respecting gauge invariance and parity
invariance produces a Chern-Simons term as a one-loop effect. Then,
in 3 dimensions, if one is going
to study a gauge theory with charged fermions, the CS term is there.

As I mentionned above, in the presence of a CS term the photon becomes
massive so that long-range correlations, necessary for Debye screening
of the monopole charged gas, disappear.
One should then expect that monopoles and anti-monopoles do not respond
anymore as a charged plasma but as a gas of "molecules" or "dipoles"
which does not exhibit Debye screening. If this were true, the Wilson loop
for external heavy charges would no longer have an area law and confinement
should be destroyed.  In particular, such a result should be welcome
in models intended to describe
the physics of high temperature superconductors where
spinless charge carriers  and neutral spin carriers should
be {\it separatedly} part of the physical spectrum (This being impossible
in a confined phase).

The deconfinement scenario was discussed by several authors \cite{Pi}-
\cite{Aff}. I will describe in this talk the approach we developed
in collaboration
with Eduardo Fradkin \cite{FS}. From our treatment, it will be clear that
the lack of gauge invariance of the fermionic determinant in a monopole
background is at the origin of deconfinement of electric charge. (Remember
that the Chern-Simons term arises as a result of violation of parity
invariance when computing the fermion determinant in a gauge-invariant
way. This means that the resulting CS term is necessarily gauge-invariant
under {\it topologically trivial} gauge transformations or in topologically
trivial gauge field backgrounds but not when topology enters in the game).

That instanton effects might be wiped out by
anomalies can be tested in other simpler models. In particular, I will
analyse in the second part of my
talk a two-dimensional model (the abelian chiral Higgs
model) in which the coupling to Weyl fermions produce a (gauge) anomaly.
Again instanton effects disappear, as we have shown in collaboration
with Marta Trobo \cite{ST}.

\section{Confinement in $CQED_3$}

Let me start by describing how Polyakov \cite{Po2} proved
confinement of electric
charge in $CQED_3$ without CS term . Consider the $SO(3)$
Georgi-Glashow model in d=3 euclidean dimensions with Lagrangian:

\beq
L = \frac{1}{4} \vec F_{\mu\nu}^2 + \frac{1}{2} D_{\mu}\vec{\phi}^2 +
V[\vec \phi^2],  \label{2}
\eeq

\beq
D_{\mu}\vec{\phi} = \partial_{\mu}\vec{\phi} + e \vec{A_{\mu}}\wedge \vec{\phi}
\label{3}
\eeq
and $V[\vec{\phi}^2]$ a symmetry breaking potential taking its minimum at
$ \vec{\phi} = \vec{\phi}_0 $. The Euler-Lagrange equations arising from
(\ref{2}) have regular solutions with finite action which are just the
static 't Hooft-Polyakov \cite{tH}-\cite{Po1} monopole solutions of the
corresponding (3+1) model. Although its exact form is not important here,
let us indicate that if we define the electromagnetic tensor
$ {\cal F}_{\mu\nu} $ associated with the residual $U(1)$ symmetry
so that:

\beq
\lim_{\vert x \vert \rightarrow \infty}{\cal F}_{\mu\nu} = \vec{\phi}\cdot
\vec{F}_{\mu\nu}, \label{4}
\eeq
then the magnetic field $B_{\mu}^{mon}$ for a monopole located at
$\vec x = \vec R$ is given by:

\beq
B^{mon}_{\mu}(\vec{x},\vec {R}) = \frac{1}{2} \epsilon^{\mu\nu\alpha}
{\cal F}^{mon}_{\nu\alpha}   \label{5}
\eeq
so that al large distances,

\beq
\lim_{x \rightarrow \infty} B^{mon}_{\mu}(\vec{x},\vec {R}) \sim
\frac{1}{2} \frac{x_{\mu} - R_{\mu}}{\vert {\vec x} - {\vec R} \vert^3}
\label{5A}
\eeq

These monopole solutions have finite action and can be taken as instantons in a
non-perturbative analysis of $CQED_3$. To this end, consider the partition
function for the model with dynamics described by Lagrangian (\ref{2}):

\beq
\int D\vec{A}_{\mu}D\vec{\phi} exp(-S[\vec{A}_{\mu},\vec{\phi}]) \label{6}
\eeq
with

\beq
S[\vec{A}_{\mu},\vec{\phi}] = \int d^3x L \label{7}
\eeq

Let us now perform a semiclassical expansion by first
considering small fluctuations around a charge-1 monopole solution:

\beq
A_{\mu} = A_{\mu}^{mon}(\vec x,\vec R) + a_{\mu} \label {8}
\eeq

\beq
\phi_{\mu} = \phi^{mon}(\vec x,\vec R) + \varphi \label{9}
\eeq
or, calling $F \equiv (A_{\mu},\phi)$, $f \equiv (a_{\mu},\varphi)$,

\beq
F = F^{mon}(\vec x,\vec R) + f. \label {10}
\eeq

One then has for action (\ref{7}) up to second order in fluctuations:

\begin{eqnarray}
S[A_{\mu},\phi] & = & S[A_{\mu}^{mon}(\vec x,\vec R),\phi^{mon}(\vec x,\vec
R)] + \int dx f(x) \underbrace{\frac{\delta S}{\delta F(x)}}_{=0}
\left.\right\vert_{F^{mon}} + \nonumber \\
& &\int dx dy f(x)\frac{\delta^{2}S}{\delta F(x)\delta F(y)}
\left.\right\vert_{F^{mon}}
f(y) \label {11}
\end{eqnarray}

or:

\beq
S[A_{\mu},\phi] = S^{mon} + \int d^3x d^3y f(x) S^{II}(x,y) f(y) \label{12}
\eeq

One has to be carefull in using (\ref{12}) due to the existence of zero-modes
related to invariances of the classical theory. In particular, associated
with translation invariance there is a direction in which the integral over
Bessel-Fourier coefficients $c_n$:

\beq
F = F^{mon} + \sum_{n} c_nf_n \label{13}
\eeq
is not gaussian. (In (\ref{13}) $f_n$ are the eigenfunctions of
the quadratic form in $S^{II}$).
One cannot then just write for the partition function measure:
\beq
DF = \prod_{n}dc_n \label{14}
\eeq
since then Z becomes infinite. Instead, one eliminates from the
sum in eq.(\ref{13}) (the product in (\ref{14})) the coefficient accompanying
the zero-mode, trading it by the collective coordinate $\vec R$ fixing the
position of the monopole. One then has instead of (\ref{14}):

\beq
DF =  N d\vec R {\prod}'dc_n \label{15}
\eeq
where the prime indicates that the zero-mode contribution has been eliminated
from the product and $N$ is a normalization constant.
In a completely analogous way one handles the problem of gauge zero-modes.
Once this is done, one is left with gaussian integrations leading to:

\beq
Z^{(1)} = N\int D\vec R exp[-S^{mon}]det^{-\frac{1}{2}}S^{II} det^{\frac{1}{2}}
\Delta_{FP} \label{16}
\eeq
with the superscript $(1)$ indicating the charge-1 monopole contribution
to the partition function and $\Delta_{FP}$ the Faddeev-Popov operator.

In order to compute the contribution to $Z$ coming from arbitrarily charged
monopoles (i.e., to include all topological sectors), we shall consider,
following ref.\cite{Po2}, a superposition
of $N$ widely separated monopoles of charge $\pm 1$ leading to a charge $n$
configuration. The radius  of each $\pm 1$ monopole is of the order
of the inverse of the vector meson mass $M_W \sim e\phi_
0$ so that if we call
${\vec R}_a$ the position of the $a$-th $\pm 1$ monopole, $a = 1,2, \dots
N$, widely separated means:

\beq
R_{ab} \equiv \vert {\vec R}_a - {\vec R}_b \vert \gg
\frac{1}{M_W} \label{17}
\eeq
Performing an expansion as that in eq.(\ref{11}) in each monopole sector
one arrives to:

\beq
S = S^{mon (N)} + S^{II} \label{17A}
\eeq
with:
\beq
S^{mon(N)} = \sum_{n_a = \pm 1} n_a^2 S^{mon} +
\frac{2\pi}{e^2} \sum_{n_a \ne n_b} \frac{n_a n_b}{R_{ab}} +
O(\frac{1}{M_W R_{ab}})  \label{18}
\eeq
the action for an $N$ monopole superposition.

With this, one can compute the contribution of all topological sectors
to the partition function. The answer is:

\beq
Z = \sum_{N, \{n_a\}} \int \prod_a d{\vec R}_a \frac{\xi ^N}{N!}
exp[-\frac{2\pi}{e^2} \sum_{n_a \ne n_b} \frac{n_a n_b}{R_{ab}}]
\label{19}
\eeq
where $\prod_a d{\vec R}_a $ is the integration measure over all
monopole locations in a given superposition,

\beq
\xi =M_W^{\frac{7}{2}} exp[-S^{mon}]
det^{-\frac{1}{2}} S^{II} det^{\frac{1}{2}} \Delta_{FP}
\label{20}
\eeq
and $\{n_a\}$ represents different superpositions of charge $\pm 1$
monopoles leading to a charge $n$ configuration. Accordingly, the $N!$ has
been included in order to avoid double counting.
Here we have used the factorization of determinants in an $N$-monopole
background into the product of determinants in a $\pm 1$ monopole
background (valid whenever condition (\ref{18}) holds.

Now $Z$ as given by eq.(\ref{19}) coincides with the partition function for
a Coulomb gas of {\it magnetically} charged particles (with charge $ n_a
= \pm 1$) interacting through a $\frac{1}{R_{ab}}$ potential. This gas exhibits
Debye screening of magnetic charge, this being in turn responsible for
confinement of electric charge. Indeed, as shown by Polyakov \cite{Po2}
the Wilson loop computed from the model with partition function (\ref{19})
exhibits an area law behavior:

\beq
\lim_{T \rightarrow \infty} <exp[i\oint A_{\mu}^3 dx^{\mu}> \sim exp[-E(R)T]
\label {21}
\eeq

\beq
E(R) = \gamma R
\label{22}
\eeq
with $R$ the distance between two external electric test-charges and $\gamma$
a constant calculable in terms of $\xi$ as given by (\ref{20}).

\section{Adding a Chern-Simons term}

Either one adds (massive) fermions or a Chern-Simons term to the Lagrangian
(\ref{2}) the  confinement scenario described above is radically changed.

To see this, let us first remind that, given the fermionic Lagrangian in
$d=3$ Euclidean dimensions,

\beq
L_F = \bar \psi (i\not\!\partial + e\not\!A + i m)\psi \equiv \bar \psi
D[A] \psi \label{23}
\eeq
the associated fermion determinant up to one-loop takes the form \cite{R}:

\beq
log det D[A] = \frac{m}{\vert m \vert} \frac{ie^2}{8\pi^2} S_{CS} +
\frac{1}{4\pi \vert m \vert} \int d^3x F_{\mu\nu}^2   ,\label{24}
\eeq
where $S_{CS}$ is the Chern-Simons action:

\beq
S_{CS} = tr \int d^3x \epsilon^{\mu\nu\alpha} (A_{\mu}\partial_{\nu}A_{\alpha}
+ \frac{2}{3} e A_{\mu}A_{\nu}A_{\alpha})
\label{25}
\eeq

As mentionned before, the emergency of this parity non-conserving term is
due to the impossibility of regularizing the fermionic determinant both
respecting gauge and space-time reflection invariances \cite{R}. As it is
well-known, the CS term is topological in the sense it does not depend on
the metric. Then, either in Minkowski or Euclidean space it appears with
an $i$ factor in the total effective action resulting from integrating out
fermions:

\beq
S_{eff} = S[{\vec A}_{\mu},\vec \phi] + i \mu S_{CS} \label{26}
\eeq
with $S[{\vec A}_{\mu},\vec \phi]$ as defined in
(\ref{7}) and $\mu = \frac{e^2}{8 \pi}$.

Solutions to the equations of motion associated
with action (\ref{26}) are complex and, what is worse, they lead to
an infinite action \cite{Pi}. Leaving aside these solutions, which are
useless in a non-perturbative calculation, we shall study the effect
of adding the Chern-Simons term when the {\it old} monopole solutions
are taken as instantons. This is the natural thing to do if one considers
the CS term as a one-loop effect arising from integration of fermions.
Now, although  $S_{CS}\left \vert _{A_{\mu}^{mon}}\right. = 0$, one has:

\beq
\frac{\delta S_{eff}}{\delta A_{\mu}} \left \vert _{A_{\mu}^{mon}}\right.
\ne 0,
\label{27}
\eeq
then, repeating the calculation in Section 2 in order to integrate
quadratic fluctuations one get, appart from the classic and quadratic
terms already present in (\ref{16}), a linear term:

\beq
S_{eff} = S^{mon} + \frac{ie^2}{\pi}\int d^3x B^{mon}_{\mu}a^{\mu} + S^{II},
\label{28}
\eeq
One can easily eliminate this linear term but then the classical term is
modified:

\beq
S_{eff} = S^{mon} + \frac{e^4}{\pi^2}\int d^3x d^3y B^{mon}_{\mu}(x)
S^{(2)\mu\nu}(x,y) B_{\nu}^{mon}(y) + {\bar S}^{II}. \label{29}
\eeq
where  the actual form of $S^{(2)\mu\nu}(x,y) $ is not important here

Again, one considers a monopole superposition as before, performs the
change of variables (\ref{10}), separates out collective coordinates
${\vec R}_a$, etc. The new term in (\ref{29}) gives an extra
contribution arising from the magnetic monopole field:

\beq
\lim_{x \rightarrow \infty}\vec{ B}^{mon} (\vec{x},\vec {R}) \sim
\frac{1}{2} \sum_{a} n_a \frac{\vec x - {\vec R}_a}{\vert {\vec x} - {\vec R_a}
 \vert^3} \label{30}
\eeq

I will skip details and just quote from Ref.\cite{FS} the relevant contribution
coming from the new term in (\ref{29}):

\beq
\frac{e^4}{\pi^2}\int d^3x d^3y B^{mon}_{\mu}(x)
S^{(2)\mu\nu}(x,y) B_{\nu}^{mon}(y)  = - \frac{e^2}{16 \pi ^2}\sum_{n_a,
n_b} n_a n_b R_{ab} \label{31}
\eeq
so that, instead of partition function (\ref{19}) one now gets:

\beq
Z  = \sum_{N, \{n_a\}} \int \prod_a d{\vec R}_a \frac{\xi ^N}{N!}
exp[-\frac{2\pi}{e^2} \sum_{n_a \ne n_b} \frac{n_a n_b}{R_{ab}}
+  \frac{e^2}{16 \pi ^2} \sum_{n_a , n_b} n_a n_b R_{ab}]
\label{32}
\eeq

The presence of the linear term in (\ref{32}) implies that confinement is
destroyed. Monopoles and antimonopoles themselves become confined due
to a linear potential, forming a gas of molecules instead of a charged plasma.
Debye screening is lost and the linear potential between electric charges
disappears.

There is an alternative way to see that monopole contribution is wiped out
from the partition function without resource of non-perturbative calculations.
I shall describe it in the next section.

\section{Integrating over all field configurations}

Let us consider for simplicity an abelian gauge theory although the $SO(3)$
case can be identically treated. The partition function for or model is:

\beq
Z = \int DA_{\mu}D\bar \psi D\psi exp[-\int d^3x F_{\mu\nu}^2 +
\int d^3x \bar \psi D[A] \psi  ,\label{32A}
\eeq
where $D[A]$ is the covariant Dirac operator for (massive) fermions. It
is important to stress that the gauge field integration in (\ref{32}) is
extends {\it over all gauge field configurations} (See any Quantum Field
Theory textbook). Usually, one makes this explicit by means of the
Faddeev-Popov technique ending with:

\beq
Z = \int \Delta_{FP}[A] DA_{\mu}\delta(F[A^{\omega}])D\omega D\bar \psi
\psi exp[-\int d^3x F_{\mu\nu}^2 + \int d^3x \bar \psi D[A] \psi
\label{33}
\eeq
Here $F[A^{\sigma}] = 0$ is the gauge fixing condition selecting one
representative  $A^{\sigma}$ over each gauge orbit and $\Delta_{FP}[A]$
is the corresponding Faddeev-Popov determinant related to the natural
metric over orbit space $\Gamma[A^{\sigma}]$ and the scale $\rho[A^{\sigma}]$
of each orbit \cite{BV}:

\beq
\Delta_{FP}[A^{\sigma}] \equiv \rho[A^{\sigma}](\Gamma[A^{\sigma}])
^{\frac{1}{2}}. \label{33A}
\eeq
Finally, $D\omega$ is the volume element on the group of gauge transformations.

Usually one eliminates the $\omega$-dependence in the integrand in (\ref{33})
by changing variables:

\beq
A \rightarrow A' = A^{-\omega} \nonumber
\eeq

\beq
\psi \rightarrow \psi' = exp[i\omega] \psi \nonumber
\eeq

\beq
\bar \psi \rightarrow \bar \psi ' = \bar \psi exp[-i \omega]
\label{34}
\eeq

If the associated Jacobian is trivial, itegration over $\omega$ factorizes.
The point is that in a monopole background, the fermionic measure changes
non-trivially under transformations (\ref{36}):

\beq
D\mu_A[\psi] \equiv exp(-S_F[A,\bar \psi,\psi])D\bar \psi D\psi =
J[A,\omega] D\mu_A[\psi ']
\label{35}
\eeq
with

\beq
J[A,\omega] = \frac{det D[A^{\omega}]}{det D[A]} \label{36}
\eeq
The reason why $J[A,\omega] \ne 1$ is the following: each one of the
determinants in (\ref{36}) contains, as we have seen, a Chern-Simons term
which is not invariant under gauge transformations in a monopole background.
Indeed:

\beq
S_{CS}[A^{\omega}] = S_{CS}[A] + \frac{e^2}{8\pi^2}\int d^3x B^{mon}_{\mu}
\partial^{\mu}\omega
\label{37}
\eeq
Integrating by parts and dropping surface terms one has:

\beq
S_{CS}[A^{\omega}] - S_{CS}[A] = -\frac{e^2}{8\pi^2}\int d^3x
\partial^{\mu}B^{mon}_{\mu} \omega
\label{38}
\eeq
Now, in this abelian example, the magnetic field of a monopole at
$\vec x = \vec R$ with magnetic charge $n$ satisfies:

\beq
\partial^{\mu}B^{mon}_{\mu} = \frac{4\pi n}{e} \delta^3(\vec x - \vec R)
\label{38A}
\eeq
(Subtleties related to the construction of abelian monopoles, in particular
concerning the appearence of Dirac strings will not be taken into account
since we have in mind the $SO(3)$ model where regular monopole solutions
without Dirac strings exists. As we shall explain, the arguments presented
in the abelian case are completely rigorous in the non-abelian one).

Then, from eqs.(\ref{37})-(\ref{38}) we have:

\beq
S_{CS}[A^{\omega}] - S_{CS}[A] = \frac{1}{2\pi} n \omega(\vec R)
\label{39}
\eeq
With this, the Jacobian (\ref{36}) becomes:

\beq
J[A,\omega] = exp[-in \omega(\vec R)]
\label{40}
\eeq
so that, when one integrates out $\omega$ all topologically non-trivial
(i.e. $n \ne 0$) sectors are wiped out from the partition function since:

\beq
\int \prod_{x} d\omega (x) exp[-in \omega(\vec R)] \propto \delta_{n,0}
\label{41}
\eeq
Even configurations with total net charge zero but consisting of a
superposition of equal number of monopoles and antimonopoles do not
contribute. As an example, consider a superposition of a $+1$ monopole
located at ${\vec R}_1$ and an antimonopole of charge $-1$ at ${\vec R}_2$.
Then, instead of (\ref{41}) one has for the $\omega$-integration:

\begin{eqnarray}
\int \prod_{x} d\omega (x) exp[-i \omega ({\vec R}_1) +i \omega({\vec R}_2)]
& = & \int d\omega ({\vec R}_1) exp[-i \omega ({\vec R}_1)] \times \nonumber
\\
& \int & d\omega ({\vec R}_2) exp[ i \omega ({\vec R}_2)] = 0 \label{42}
\end{eqnarray}

Thus we see that the partition function only picks a
contribution from the no-monopole sector. Since confinement was precisely
produced by monopole contributions, we see that when fermions (or a
Chern-Simons
term) is included, electric charges are no more confined by a linear potential.
The only modification to the arguments above in the non-Abelian case arises
from the fact $\omega$ takes values in the Lie algebra of the
gauge group. For monopoles such that the residual $U(1)$ symmetry corresponds
to the $3^{rd}$ $SO(3)$ direction, this leads to a Jacobian of the form:

\beq
J[A,\omega] = exp[-in \omega^3(\vec R)]
\label{43}
\eeq
{}From this results, the conclusions reached for the abelian case trivially
extend to the non-abelian model.

\section{A two-dimensional model}

As we stated in the Introduction, in any model where classical invariances
might be spoiled at the quantum level one should revise instanton calculations
since topologically non-trivial sectors might be wiped out from the correctly
gauge-fixed partition function. In the precedent Section we showed how
parity anomaly, through the emergence of a Chern-Simons term when computing
the fermionic path-integral, eliminates monopole contributions so that
confinement of electric charge is destroyed.

There is a natural candidate to analyse whether the same phenomenon
happens: the 2-dimensional abelian Higgs model coupled to {\it chiral}
fermions. As it is well-known, the abelian Higgs model, with action:

\beq
S_H = \int d^2x \left( -\frac{1}{4} F_{\mu\nu}^2 + \frac{1}{2} \vert
(\partial_{\mu}
-ie A_{\mu}) \phi \vert^2 + V[\vert \phi \vert^2]\right) \label{50}
\eeq
has vortex solutions \cite{NO} which can be taken as instantons in
2-dimensional Euclidean space. (Here, $\phi$ is a complex
scalar and $V[\vert \phi \vert^2]$ a symmetry breaking potential having
its minimun at $\phi = \phi_0$).
Precisely when one takes into account this
instantons in a non-perturbative analysis of the model, one discovers
screening of (fractional) electric charge due to Debye screening produced
in the vortex plasma \cite{CDG},\cite{RU},\cite{FAS}. More recently, the
model has received much attention
since it provides a laboratory to analyse if instanton effects can lead
at high energy to  fermion number violation, a phenomenon of main
revelance in the analysis of the Standard Model \cite{KR}-\cite{McL}.

What happens if one adds chiral (say left-handed) fermions to the
model? Of course, the corresponding fermionic current is not conserved
due to the presence of the anomaly. Nevertheless, it is by now accepted that
the so-called anomalous models can be consistently quantized
if one correctly takes into account the gauge degrees of freedom
\cite{JR},\cite{FdS},\cite{BSV},\cite{HT}. Of course many questions about
renormalizability and unitarity of the resulting quantum theory remain
to be investigated but in the particular case of 2-dimensional models,
this problems do not exists \cite{JR} so that it is a sensible question
to analyse whether integration over gauge degrees of freedon wipes out
instanton effects as it  does in the 3-dimensional model described
in Section 4.

To answer this question, let us add to the abelian Higgs action (\ref{50})
left-handed fermions with action:

\beq
S_F = \int d^2x \bar \psi D[A]\psi \label{51}
\eeq

\beq
D[A] = (i\not\!\partial + e\not\!A)\frac{(1 + \gamma_5)}{2}  \label{52A}
\eeq
and consider the partition function:

\beq
Z = \int D\phi DA_{\mu} D\bar \psi D\psi exp[-(S_H +S_F)] \label {52B}
\eeq
As before, we write the $A_{\mu}$ measure \`a la  Faddeev-Popov, ending
up with:

\beq
Z = \int D\phi DA_{\mu} \Delta[A] \delta(F[A])
J[A,\omega] D\omega exp[-(S_H + S_F)].
\label{53}
\eeq
Again, we have to determine whether the Jacobian $J[A,\omega]$,

\beq
J[A,\omega] = \frac{det D[A^{\omega}]}{det D[A]} \label{54}
\eeq
is trivial or not. To see this, let us note that each determinant in
(\ref{54}) is not defined since the Dirac operator (\ref{51}) does not have
an eigenvalue problem (it maps negative chirality fermions into positive
chirality ones). The chirality-flip problem is usually overcome by defining:

\beq
det D[A] \equiv \det \hat{D}[A]\vert_{reg}, \label{E}
\eeq
where

\beq
\hat{D}[A] = D[A] +i\not\!\partial\frac{1}{2}(1+\gamma_5) \label{F}
\eeq
and $\vert_{reg}$ means that an appropriate regularization scheme has
been adopted to make sense from the (originally unbounded) product of
eigenvalues defining the determinant. Now, the addition of free right-handed
fermions solves the chirality-flip problem  but creates
a new one: under a gauge transformation
\beq
A_{\mu} \to A_{\mu}^{\theta} = A_{\mu} + \frac{1}{e} \partial_{\mu}\theta
,\label{G}
\eeq
the Dirac operator $\hat{D}[A]$ does not transform  as a
covariant derivative and then one has in general

\beq
\det \hat{D}[A^\theta]\vert_{reg} \ne \det \hat{D}[A]\vert_{reg}. \label{H}
\eeq

This means that $J[A,\omega]$ is, in principle, non-trivial. In fact, it
is easy to find that \cite{BSV}-\cite{HT}:

\beq
log J[A,\omega] = -\frac{1}{4\pi}\int d^2x[\frac{(a-1)}{2}\partial_{\mu}\omega
\partial_{\mu}\omega +e(a-1)A_{\mu}\partial_{\mu}\omega +e\epsilon_{\mu\nu}
A_{\mu}\partial_{\nu}\omega]. \label{XXX}
\eeq
Here $a$ is a real parameter which takes into account regularization
ambiguities which arise when computing gauge non-invariant determinants
\cite{JR}. This non-trivial Jacobian induces a Wess-Zumino term which absorbs
the anomaly rendering the theory gauge-invariant. Indeed, if
we define the fermionic effective action $S_{eff}$ :

\beq
exp(-S_{eff}[A]) \equiv \int D\omega D\bar\psi D\psi
J[A,\omega] exp(-S_{F}[A,\bar\psi,\psi]),     \label{R}
\eeq
one can easily verify, using the one-cocycle \cite{FS} condition
satisfied by the Jacobian $J[A,\omega]$,

\beq
log J[A,\theta + \omega] = log J[A,\theta] + log J[A^{\theta},\omega],
\label{S}
\eeq
that $S_{eff}$ is gauge-invariant:

\beq
S_{eff}[A^{\theta}] = S_{eff}[A]. \label{T}
\eeq
This result implies that the fermionic current, defined as $
{\delta S_{eff}}/{\delta A_{\mu}}$ is conserved. We shall then take as the
partition function for the chiral Abelian Higgs model:

\beq
Z = \int D\phi DA_{\mu} \Delta[A] \delta(F[A])
J[A,\omega] D\omega exp[-(S_H[A,\phi] + S_{eff}[A])] \label{TR}
\eeq

As we stated above, eq.(\ref{T}) guarantees conservation of
the fermionic current and this indicates that the theory with
partition function $Z$ given by eq.(\ref{TR}) should be consistently
quantized. Of course, in general, unitarity and renormalizability must be
investigated. For the two-dimensional chiral Schwinger model and its
non-abelian extension \cite{virginia}, it has been shown
(see \cite{fas} and references therein)
that the proposal of refs.\cite{BSV}-\cite{HT} leads to a consistent,
unitary  and Lorentz invariant quantum theory both in the path-integral and
canonical quantization approaches \cite{abdalla}.
As we shall see bellow, the same holds for the chiral Higgs model
in two dimensions.

What about instanton contributions to the theory with partition function
(\ref{TR}) ? As we stated above, the Abelian Higgs model has vortex-like
solutions \cite{NO} which can be taken as instantons in the computation
of non-perturbative effects \cite{CDG}-\cite{FAS}. Asymptotically,
a Nielsen-Olesen vortex configuration takes the form:

\beq
\lim_{r \to \infty} A_{\mu}^{vortex}(r,\varphi) =
\frac{n}{e}\partial_{\mu}\varphi \label{U}
\eeq

\beq
\lim_{r \to \infty} \phi^{vortex}(r,\varphi) = exp(in\varphi) \phi_{0},
\label{V}
\eeq
Such
a configuration carries $n$ units of magnetic flux (i.e., it has a topological
charge equal to $n$):

\beq
\frac{e}{2\pi} \oint A_{\mu}^{(vortex)}dx^{\mu} = n. \label {W}
\eeq

The path-integral can then be performed in each
topological sector (A configuration
satisfying (\ref{U})-(\ref{V}) being a representative in each sector) .Using
a superscript "$n$" to indicate the topological sector to which Higgs and
gauge fields belong, the partition function will then be written in the form:

\beq
Z = \sum_{n} \int D\phi^{(n)} DA_{\mu}^{(n)} \Delta[A^{(n)}] \delta(F[A^{(n)}])
exp(-S_{Higgs}[\phi^{(n)},A^{(n)}]-S_{eff}[A^{(n)}]).  \label{X}
\eeq

A comment on zero-modes of the
Dirac operator and the fermionic determinant appearing in the effective
action (\ref{R}) is here in order. We know that the operator $\not\! D[A^{(n)}$
acting on{\sl Dirac} fermions has $\vert n \vert$ square integrable zero-modes
For $n > 0$ ($n < 0$) these zero-modes are right handed (left-handed)
\cite {jacros}.
Hence, the corresponding regularized fermion determinant vanishes for all
$n \ne 0$. This automatically ensures that  only the
$n=0$ sector contributes to $Z$ for Dirac fermions\cite {cabvon}. On the
contrary, the operator $\hat D[A^{(n)}]$ has only left-handed zero-modes since
in the right-handed sector it coincides with the free Dirac oparator which
does not have normalizable zero modes).
Then $det \hat D[A^{(n)}]$ vanishes only in the case $n<0$. Consequently
$Z$ reduces to:

\beq
Z = \sum_{n>0} \int D\phi^{(n)} DA_{\mu}^{(n)} \Delta[A^{(n)}]
\delta(F[A^{(n)}]) exp(-S_{H}[\phi^{(n)},A^{(n)})-S_{eff}[A^{(n)}]).
\label{XX}
\eeq

We have now to use the explicit result (\ref{XXX})
for the Jacobian. Of course, since we are working in topologically non-trivial
sectors, we must not drop surface terms when performing the
$\omega$-integral in $S_{eff}$. Indeed, gauge parameters $\omega$
not vanishing at infinity are compatible with any imposed boundary condition
at $r=\infty$ since for example $\omega = 2\pi$ is equivalent to $\omega
= 0$. This gives, for the second term in the argument of the
exponential in (\ref{X}):

\beq
e(a-1)\int d^x A_{\mu}^{(n)}\partial_{\mu}\omega = e(a-1)\int d^2x
\omega \partial_{\mu} A_{\mu}^{(n)} + e(a-1)\oint dx^{\mu}\omega A_{\mu}^{(n)}
\label {Y}
\eeq
or

\beq
e(a-1)\int d^x  A_{\mu}^{(n)}\partial_{\mu}\omega = e(a-1)\int d^2x
\omega \partial_{\mu} A_{\mu}^{(n)} + e(a-1)2\pi n\omega_{\infty}
\label {Z}
\eeq
where we have used (\ref{W}) and called $\omega_{\infty}$ the value of
$\omega$ at infinity. We then see that for $n \ne 0$ surface terms do
contribute in a non-trivial way. With all this, we get after integrating over
$\omega$:

\begin{eqnarray}
exp(-S_{eff}[A^{(n)}]) = \int D\bar\psi D\psi
exp[-(S_{F}[A^{(n)},\bar\psi,\psi] + \nonumber\\
+\frac{e^2}{8\pi(a-1)} \int d^2x A_{\mu} A_{\mu} + F^{(n)})].  \label{A1}
\end{eqnarray}
The second term in the argument of the exponential is the usual one obtained
(in the Lorentz gauge) after integration over $\omega$ when non-trivial
topological sectors are not taken into account \cite{BSV}-\cite{HT}.
The third term precisely corresponds to the border contribution and is given
by:

\beq
F^{(n)} = {\cal N}\lim_{R \to \infty} F^{(n)}(R), \label {B1}
\eeq
where

\beq
F^{(n)}(R) = \frac{n^2a^2}{8(a-1)}logR, \label {C1}
\eeq
and ${\cal N}$ is a constant.

Then, after taking the limit $R \to \infty$ we have:

\beq
exp(-S_{eff}[A^{(n)}]) = exp(-S_{eff}[A^{(0)}]) \delta_{n,0} \label{D1}
\eeq
and hence the partition function Z (eq.\ref{XX}) only picks contribution
from the $n=0$ sector:

\beq
Z = \int D\phi^{(0)} DA_{\mu}^{(0)} \Delta[A^{(0)}] \delta(F[A^{(0)}])
exp(-S_{H}[\phi^{(0)},A^{(0)}]-S_{eff}[A^{(0)}]).  \label{D2}
\eeq

As announced, the chiral Higgs model, though
anomalous can then be consistently quantized and only the $n=0$ sector
contributes to the partition function. The value of $S_{eff}[A]$
can be easily evaluated from (\ref{R}). The answer (in the Lorentz gauge)
is \cite{BSV}-\cite{HT}:

\beq
S_{eff}[A] = \frac{a^2}{8\pi (a-1)}\int d^2x A_{\mu}A_{\mu}                  .
 \label{D3}
\eeq

Inserting this value in (\ref{D2}) we see that the result coincides with that
corresponding to the Abelian Higgs model with {\sl Dirac} fermions except
for the fact that the vector meson mass $m_v$ is given by

\beq
m_v^2 =
e^2(\vert\phi_0\vert^2 + {a^2}/{4\pi}(a-1)). \label{ultima}
\eeq
Then, $Z$ in (\ref{D2}) defines a unitary, positive model for any value of the
parameter $a$ such that $m_v^2>0$. The existence of a whole range
of the undetermined parameter $a$ for which the model is consistent,
Lorentz invariant and unitary with a vector meson mass which is not fixed
by gauge invariance (as it happens for models with Dirac fermions) is typical
of anomalous gauge theories  at least in $d=2$ dimensions.
One can think that $Z$ defines a family of quantum theories and that the
ultimate value for $a$ should be determined by physical considerations (see
ref.\cite{jactei} for a discussion on these facts).

In summary, we have shown in this Section that the chiral Abelian Higgs model
in two dimensions, though anomalous, can be consistently quantized following
the proposal of refs.\cite{BSV}-\cite{HT}. The anomaly is cancelled
by a "Wess-Zumino" term, as suggested in \cite{FS} and as a result,
instanton contributions are eliminated from the partition function defining
the quantum theory. The procedure can be straightforwardly generalized to
non-Abelian models and one can also envisage the analysis of four dimensional
Higgs models. However, in this last case the issues of unitarity and
renormalizability should be carefully investigated.

\section{Questions}

\noindent \underline{N.Bralic}: {\it You used an abelian example to show
that integrating over
all field configurations eliminates monopole contributions to the partition
function. In your proof, it was crucial that
$\partial_{\mu}B_{\mu} = \delta ^3(\vec x - \vec R)$. Can you explain how
does your argument work for the non-abelian monopole, for which you do not
have such a relation?  }

\vspace{0.2in}

\noindent Answer: The 't Hooft-Polyakov (charge 1) monopole solution reads
\cite{tH}-\cite{Po1}:

\beq
\phi^a = {\hat x}^a f(r) \label{I}
\eeq

\beq
A_{\mu}^a = \epsilon_{a \mu i} {\hat x}^i \frac{(1-K(r))}{r}
\label{II}
\eeq
with ${\hat x}^a \equiv {x^a}/{r}$, $f(0) = 0$, $f(\infty) = \phi_0$,
$K(0) = 1$, $K(\infty)= 0$. The corresponding magnetic field is
(see eq.(\ref{5})):

\beq
B_{\mu} = \frac{{\hat x}_{\mu}}{r^2}(1 - K^2) \label{III}
\eeq
so that:

\beq
\Phi = \int B_{\mu}dS_{\mu} = 4\pi \label{IV}
\eeq

Let us consider a family of gauge transformations $g$ of the form:

\beq
g = exp[\frac{i}{2} \omega(r) \sigma^a {\hat x}^a] \label{VV}
\eeq
such that $\omega(0) = 0$ (in order to avoid singularities) and
$\omega(\infty) = \omega$ (with $\omega$ a non-zero constant).
Under such a transformation, the monopole configuration (\ref{II})
becomes:

\beq
\left(A_{\mu}^{a}\right)^g = \epsilon_{a \mu i} {\hat x}^i \frac{(1-K
 cos\omega(r))}{r}
+ (\delta_{a \mu} - {\hat x}_{\mu}{\hat x}_a)\frac{K sin\omega(r)}{r}
+ {\hat x}_{\mu}{\hat x}_a \frac{d\omega(r)}{dr}
\label{VI}
\eeq
so that asymptotically one has:

\beq
\left(A_{\mu}^{a}\right)^g = \epsilon_{a \mu i} \frac{{\hat x}^i}{r} +
{\hat x}_{\mu}{\hat x}_a \frac{d\omega(r)}{dr}  \label{VII}
\eeq

Let us consider how the Chern-Simons action changes under such a class of gauge
transformations:

\beq
\delta S_{CS}[A]  = \frac{2}{e^2} \epsilon_{\mu \nu \alpha} tr[
\frac{1}{3} \int d^3x (\partial_{\mu}g g^{-1})(\partial_{\nu}g g^{-1})
(\partial_{\alpha}g g^{-1}) + \int dS_{\mu}A_{\nu}\partial_{\alpha}g g^{-1}]
\label{VIII}
\eeq

or

\beq
\delta S_{CS}[A] = \frac{16 \pi^2}{e^2} n[g] + \frac{2}{e^2}
\epsilon_{\mu \nu \alpha} tr \int dS_{\mu}A_{\nu}\partial_{\alpha}g g^{-1}
\label{IX}
\eeq
with $n[g]$ the winding number of the $g$-transformation (note that with
$\mu = \frac{e^2}{8\pi}$ as in eq.(\ref{26}) $exp(-S_{CS})$ is not affected
by the first term in the r.h.s. of eq.(\ref{IX})). Let us consider
the second term in the r.h.s. of eq.(\ref{26}). An explicit calculation
using the form (\ref{VI}) for the monopole configuration gives:

\beq
\delta S_{CS} = \frac{16 \pi^2}{e^2} n[g] + \frac{2}{e^2}  \int
d^3x \frac{1}{r^2} \frac{d\omega}{dr}
\label{XAX}
\eeq
or

\beq
\delta S_{CS} = \frac{16 \pi^2}{e^2} n[g] + \frac{8\pi}{e^2}\omega \label{XI}
\eeq
Now, when integrating out $\omega(x)$ in (\ref{33}) one has to include
this family of gauge transformations with $ \omega \in [0,2\pi]$, so
that there is an integral of the form:

\beq
\int_{0}^{2\pi} d\omega exp(i\omega) = 0 \label{XII}
\eeq
which wipes out the charge-1 monopole contribution from $Z$. Similar
arguments hold for charge-$n$ sectors whenever $n \ne 0$.

\vspace{.4in}

\noindent \underline{H.Banerjee}: {\it You stated that the
Chern-Simons action arised
as a one-loop effect when computing the (3-dimensional) fermion determinant.
This result crucially depends on the regularization scheme you adopt. You
may obtain the Chern-Simons term using Pauli-Villars method but other
prescriptions do not give parity-violating results, in particular
when the fermion mass is strictly zero.}

\vspace{.2in}

\noindent \underline{Answer}: Although originally the Chern-Simons term
was obtained using
the Pauli-Villars method for regularizing the fermion determinant, one
can adopt alternative methods and still obtain the same result. In
particular, in Ref.\cite{GMSS} we have employed the well-honored
$\zeta$-function method showing that a carefull application of Seeley's
technique \cite{See} {\bf does lead to a Chern-Simons term even when
the fermion mass is strictly zero}, thus contradicting the results
in \cite{BBB}. Even the sign ambiguity in front of the CS action
is reobtained using $\zeta$-function
as a result of the choice of upper or lower half-plane when computing
the finite part of the $K_{-1}(x,x,D)$ kernel without the necessity
of introducing a fermion mass (see \cite{GMSS}).

\vspace{.4in}

\noindent \underline{C.Teitelboim}:{it There are many young people
in the audience and
we are morally responsible for them. You have discussed an anomalous
gauge theory as if it could be consistently quantized but it should be
stressed that anomalous theories are not unitary and hence inconsistent.
Can you comment on this?}

\noindent{Answer}: The model I have discussed (the Abelian chiral Higgs model)
is a {\bf two-dimensional} model. As in the chiral Schwinger Model
case \cite{JR}, one can show that it can be consistently
quantized in a unitary and
Lorentz invariant way \cite{ST}. Of course the issue of unitarity
and renormalizability in {\bf four-dimensional} anomalous gauge theories
using the approach described in Section 5 is not clear and as you said
it touches moral aspects which I prefer not to discuss.

\vspace{.5in}

\noindent \underline{Acknowledgements}: I wish to thank Claudio Teitelboim
and Jorge Zanelli for inviting me once more to the Santiago Meeting on
Quantum Mechanics of Fundamental Systems. I also wish to thank
the Physics Department of the University of
Illinois at Urbana Champaign for kind hospitality during the completion of this
work.

This work was supported in part by the International Cooperative Program
NSF-CONICET, through the grant NSF-INT 8902032 and CONICET funds.

\end{document}